\documentclass[12pt]{article}
\usepackage{amsmath}
\usepackage{graphicx}
\usepackage{tikz}
\usepackage{amsfonts}
\usepackage{amsmath}
\usepackage{amssymb}
\usepackage{bm}
\usepackage{geometry}
\usepackage{setspace}
\usepackage{graphicx}
\usepackage{lscape}
\usepackage{verbatim}
\usepackage{natbib}
\usepackage{xcolor}
\usepackage{tabularx}
\usepackage{ragged2e}
\definecolor{winered}{rgb}{0.5,0,0}
\usepackage[bookmarks=true, bookmarksnumbered=true, allbordercolors={1 1 1}]{hyperref}
\hypersetup{
  colorlinks   = true, 
  urlcolor     = blue, 
  linkcolor    = blue, 
  citecolor    = winered,
}
\usepackage[open=true, numbered=true]{bookmark}
\usepackage{float}
\usepackage{fancyhdr}
\usepackage{appendix}
\usepackage{scalefnt}
\usepackage{afterpage}
\usepackage[left]{lineno}
\usepackage{caption}
\usepackage{subcaption}
\usepackage{varioref}
\usepackage{threeparttable}
\usepackage{booktabs}

\usepackage{cleveref}
\usepackage{rotating}

\makeatletter
\def\blfootnote{\xdef\@thefnmark{}\@footnotetext}
\makeatother

\emergencystretch=\maxdimen
\begin{document}
\title{\LARGE{Climate change: across time and frequencies}}
 \author{
Luis Aguiar-Conraria \\ \emph{University of Minho and NIPE}\\
\and 
Vasco J. Gabriel{\small \thanks{Corresponding author. \textit{E-mail address}:
vgabriel@uvic.ca}} \\ \emph{University of Victoria and NIPE}\\
\and Luis F. Martins \\ \emph{ISCTE -- Instituto Universit\'{a}rio de Lisboa and CIMS}\\
\and Anthoulla Phella \\ \emph{University of Glasgow}}
\date{March 2025}

\maketitle
\begin{abstract}
\noindent We use continuous wavelet tools to characterize the dynamics of climate change across time and frequencies. This approach allows us to capture the changing patterns in the relationship between global mean temperature anomalies and climate forcings. Using historical data from 1850 to 2022, we find that greenhouse gases, and CO$_2$ in particular, play a significant role in driving the very low frequency trending behaviour in temperatures, even after controlling for the effects of natural forcings. At shorter frequencies, the effect of forcings on temperatures switches on and off, most likely because of complex feedback mechanisms in Earth's climate system. 

\bigskip

\bigskip \medskip

\end{abstract}
\thispagestyle{empty} 

\newpage
\onehalfspace

\section{Introduction}

The statistical analysis of climate data has played an important role in informing the debate around the causes of climate change, as it
is critical to document how the patterns of the relationship between climate forcings and temperatures has been changing over time. Moreover, it is important to understand how forcings affect temperatures at different time horizons, i.e., what drives the dynamics of temperatures across distinct frequencies. This paper contributes to this debate by emphasizing a time-frequency domain approach through the use of wavelet tools. Wavelet analysis allows us to capture the changing patterns in the relationship between global temperatures and radiative forcings, both natural (such as solar activity) and anthropogenic (such as greenhouse gases, GHG), thus enhancing our ability to deploy different policy interventions. 

An important advantage of our proposed method is that the specification of the relationship between forcings and temperature changes is essentially nonparametric and allowed to vary over time. In particular, we will make use of the Continuous Wavelet Transform, which has recently gained traction as a valuable instrument in econometric analysis. Its time-frequency domain approach enables the simultaneous detection of time-varying patterns, while also capturing variations across frequencies. Additionally, due to its localized nature, wavelet analysis is well-suited for handling non-stationary data and nonlinear relationships. 

The use of wavelet methods is particularly appropriate in this context given the well-documented natural variability in the Earth's climate system, exhibiting substantial variations in physical feedback effects, carbon cycle mechanisms, ocean heat absorption, and effective climate sensitivity (see \citealp{MacDougallFriedlingstein2015}). These factors may give rise to nonlinearities and delayed temperature responses (\citealp{Gilletetall2013}), which in turn affect the connection between anthropogenic CO$_2$ emissions and global temperature shifts. 

In climate science, seminal contributions such as those by \cite{torrence1998practical} laid the groundwork for applying wavelet transforms to study critical climate phenomena. Building on this foundation, \cite{grinsted2004application} introduced cross-wavelet transforms and wavelet coherence, offering a robust framework for examining interactions between climate variables, such as sea surface temperatures and atmospheric pressure (see also \citealp{Mooreetal2007}, \citealp{Caccamoetal2016} and \citealp{MagazuCaccamo2018}). More recently, \cite{uddin2020wavelet} advanced the field by employing wavelet-based multiscale analysis to examine the linkages between climate change and economic growth, shedding light on how these factors interact across different time horizons.

Our approach is statistical and data-driven in nature, in contrast with the conventional practice in climate science, where models encapsulate the physical understanding of the climate system and are then simulated to produce projections.
Nevertheless, our paper relates to the literature on climate sensitivity and the transient climate response to cumulative CO$_2$ emissions (TCRE, see \citealp{Matthewsetal2009} and \citealp{Matthewsetal2018}, for example). Several studies attempt to quantify the uncertainty surrounding the timing and magnitude of warming (see \citealp{RickeCaldeira2014} or, more recently, \citealp{SpaffordmacDougall2020}), with some employing observational data to ``constrain'' these uncertainties (\citealp{Matthewsetal2009}, \citealp{Gilletetall2013}). Nevertheless, to the best of our knowledge, this literature has not explored the time-varying (conditional) effects of climate forcings across different frequencies of the temperature distribution.

The key takeaways from our analysis are the following:
\begin{itemize}
\item The relationship between temperatures and the different forcings displays considerable time variation and is heterogeneous across frequencies.
\item The relationship between temperatures and GHG/CO$_2$ is intermittently coherent at shorter frequencies but shows a stable joint trend at the lowest frequency, with notable cycles influenced by solar radiation at the 10-11 and 20-year cycles.
\item We document an increasing GHG/CO$_2$ effect on temperature acceleration, especially in recent decades.
\item The relationship between temperatures and GHG/CO$_2$, with shorter cycles showing switching correlations, is not significantly influenced by sunspots.
\end{itemize}

These results underscore the significance and intensification of anthropogenic emissions in driving temperature variations during this time frame, even after factoring in the impacts of other natural forcings. Our findings support the need for both adaptation and mitigation strategies. Indeed, the significant long-term cycles in temperature and GHG dynamics highlight the need for policies that address sustained trends rather than short-term fluctuations. This supports the implementation of long-term strategies for reducing GHG emissions to mitigate long-term warming.

Our findings are in line with a significant body of the climate econometrics literature providing ample evidence that the relationship between temperatures and climate forcings may have changed over time. In particular, our results are consistent with the presence of a ``hiatus'' in global warming, observed between 1998 and 2013, during which the warming trend temporarily slowed down while GHG levels continued to rise. Several studies offer various explanations for this phenomenon, as seen in \cite{Schmidtetal2014}, \citet{pretis_testing_2015}, \cite{Medhaugetal2017} and \citet{miller_dating_2020}, for instance.

Moreover, our results confirm the findings in \citet{agliardi_relationship_2019} and \cite{Phellaetal2024}. Indeed, \cite{agliardi_relationship_2019} find that the correlation between temperatures and GHG exhibits considerable cyclicality over their period of analysis (1870 to 2017), mostly displaying a positive relationship, but with several periods where correlations are persistently negative. On the other hand, \cite{Phellaetal2024}
find considerable variation over time in the relationship between temperatures and its drivers, and that these effects may be heterogeneous across different quantiles. These authors highlight the increasing influence of numerous anthropogenic forcings on temperatures, which exhibit an asymmetric and heightened impact on temperature anomalies' extremes; this is similar to our findings regarding the latter part of our period of analysis.

The paper is organized as follows. Section 2 provides an overview of the wavelet tools employed in the paper. In Section 3, we provide a brief description of the data and carry out the empirical analysis, studying variation in and across frequencies of temperatures and a set of climate forcings. Section 4 concludes. 

\section{Methodology: the Continuous Wavelet Transform}\label{meth}
Identifying key cyclical components in a time series typically involves applying the Fourier transform, which decomposes the series into individual frequencies. While the Fourier transform offers an alternative representation of the time series in terms of frequency, it loses time-related information, preventing the observation of frequency changes over time. As a result, Fourier analysis is only appropriate for stationary time series. A time-frequency or time-scale representation is necessary for non-stationary series, which are more common in climate data. Wavelet analysis provides an efficient solution by estimating frequency/scale characteristics as a function of time. The importance of wavelet analysis was acknowledged in 2017 when French mathematician Yves Meyer received the Abel Prize, often referred to as the Nobel Prize of mathematics, for his groundbreaking contributions to the development of wavelet theory.

This section briefly describes the continuous wavelet tools used in our analysis and explains how to interpret them. In the Appendix, there is a self-contained technical summary of our tools. For more mathematical details, we refer readers to \cite{AguiarConrariaetal2014} and \cite{AguiarConrariaetal2018}. For a broader view of digital signal processing and spectral analysis, including the Continuous Wavelet Transforms (CWT), \cite{Alessio2015} serves as an excellent guide.
\subsection{The Wavelet Power Spectrum}\label{wps}
The Wavelet Power Spectrum (WPS) measures the local variance distribution of a time series in the time-frequency domain. Rather than providing a single value representing the variance, WPS generates a matrix of values indicating the estimated variance for each moment in time and each frequency. Similar to spectral analysis, WPS allows us to identify the most significant frequencies that account for the overall variance of the time series. Additionally, wavelet analysis provides insight into when these frequency contributions are most prominent.

Consider, for example, the lower middle graph in Figure~\ref{Figure: RR_1850_Temp_GHG_Solar_WPS}, where we can observe the WPS of Solar radiance. Blue, a cold colour, dominates, indicating low volatility, while red, a warm colour, becomes prominent at the frequency corresponding to a 10-11 year cycle. This predominance persists throughout the entire sample. The implication is that this variable exhibits a fairly regular cycle with the indicated periodicity, as shown on the y-axis.

While the Discrete Wavelet Transform (DWT) provides a simple way of capturing essential time-frequency features, the CWT allows for more flexibility and offers a more refined picture of the time-frequency patterns in the data. Nevertheless, the CWT should still be discretized, and as such the wavelet power becomes a matrix, visualized typically as a heat map -- the same representation is employed for multiple and partial coherencies, as described next. This can be computationally intensive for long time series or large datasets, which is not the case in our application. It should also be noted that computing the transform at the beginning and end of the series introduces missing values, necessitating artificial prescription, leading to what are known as edge effects. The area in the time-frequency plane where the CWT is influenced by these edge effects is termed the cone-of-influence (COI, delineated by a parabola-like black line in the figures below), so we will abstain from interpreting results within this region.

\subsection{Wavelet (Partial) Coherency and (Partial) Phase Difference}\label{pwc}
In most applications, detecting and quantifying relationships between two or more time series is crucial. As \cite[p. 681]{Priestley1992} noted in the context of Fourier analysis: “(...) the whole `apparatus’ of multivariate linear regression theory can be taken over (almost unchanged) and applied to the study of multivariate spectral relationships. In particular, the ideas of ‘multiple correlation’ and ‘partial correlation’ (...) have immediate analogues in the frequency domain, where they become ‘multiple coherency’ and ‘partial coherency’.”

In the wavelet framework, wavelet coherency corresponds to Fourier coherence, and the concepts of wavelet multiple coherency and wavelet partial coherency are natural extensions of the respective concepts of multiple and partial correlation in the time domain, or (Fourier) multiple and partial coherency in the frequency domain, to the time-frequency plane.

When the wavelet is complex-valued, the wavelet transform is also complex. It can be decomposed into its real and imaginary components or represented in terms of its amplitude and phase (or phase angle). The phase angle indicates the position of the series within its cycle. Similar information can be obtained from wavelet (partial) coherency. However, in this case, since we are comparing two variables, we focus on the phase difference (the phase lead of series $y$ over series $x$). This phase difference provides insights into the delay or synchronization between the oscillations of the variables, as can be seen in Table \ref{tab1}. The partial phase difference is analogous but controls for the influence of third variables.
\bigskip
\begin{table}[htbp]
\begin{center}
\caption{Interpretation of the wavelet phase-difference between $y$ and $x$}
\label{tab1}
\begin{tabular}
{rcllll}
\hline\hline\\
$\pi $ & $\longleftrightarrow $ & $ \frac{\pi}{2}$ & & & Out-of-phase: $y$ lags\\[2ex]

$\frac{\pi}{2} $&$ \longleftrightarrow $&$ 0 $ & & & In-phase: $y$ leads\\[2ex]

$0 $&$ \longleftrightarrow $&$ -\frac{\pi}{2}$ & &  & In-phase: $y$ lags\\[2ex]

$-\frac{\pi}{2} $&$ \longleftrightarrow $&$ -\pi$ & & & Out-of-phase: $y$ leads\\[1ex]
\hline
\end{tabular}
\end{center}
\end{table}

\subsection{Wavelet (Partial) gain}\label{gain}

The concept of wavelet gain is analogous to Fourier gain. Typically, Fourier gain is understood as the modulus of the regression coefficient of $y $ on $x$ at a given frequency (e.g., \citealp{engle1976}). \cite{mandler2014} extended this interpretation to wavelet gain, applying it to specific moments in time and at particular frequencies. \cite{AguiarConrariaetal2018} further generalized this concept to encompass $n $ variables, enabling the estimation of both time-varying and frequency-varying coefficients.

Interpreting wavelet gain as a regression coefficient requires caution. Our approach is the wavelet counterpart of what \cite{engle1976} described: ``The regression coefficient is just the gain if there is no time lag between the independent and dependent variables. If there is a time lag, the gain can be interpreted as the regression coefficient if the series were lagged just the right amount to eliminate any phase shift, and the phase is the angle by which they would have to be shifted." This interpretation offers a significant advantage, though it introduces complexity to the analysis.

The primary advantage is that our regressions inherently account for variations in time lags between the variables of interest across both time and frequency domains. Thus, the model automatically adjusts the time lag specifications, avoiding the misspecification that would arise in a traditional regression when time lags change. The added complexity lies in the need to complement the analysis of wavelet gain with that of phase difference, ensuring that we identify potential time shifts and accurately interpret the estimated coefficients across different times and frequencies.

\section{Empirical Analysis}\label{empanal}

\subsection{Data Description}\label{data}

We follow the convention in the literature of working with global temperature anomalies relative to the 1986-2005 base period, which casts into sharp relief ongoing increases in temperatures in recent decades. Data for global temperatures (anomalies relative to the 1986-2005 base period) comes from Berkeley Earth, which averages raw gridded temperatures and bias-corrected station data.\footnote{We experimented with both land-only temperatures (which runs from 1750 to present), as well as averaged land and HadSST (Hadley Centre Sea Surface Temperature dataset) ocean global temperatures, starting in 1850. Results are qualitatively similar, here we report estimates based on the former.} In terms of forcing variables, we consider Solar Radiance (sunspot numbers from the Royal Observatory of Belgium), and radiative forcings (in Watts per square metre, $W/m^2$) of GHG (section 3.2), CO$_2$, natural and anthropogenic aerosols (section 3.3). All data comes from \cite{miller_cmip6_2021} (see also \citealp{Meinshausenetal2017}).\footnote{We also considered alternative formulas to approximate radiative forcings, such as \cite{Etminanetal2016}, but results are qualitatively similar.} The series are plotted in the relevant Figures in the next section. The standout features are the following: temperatures, GHG and CO$_2$ display a strong trend, which accelerates after the 1960s, sunspots exhibit a fairly regular 10-11 year cycle (although of varying amplitude), aerosols are stationary but with significant negative outliers, while anthropogenic aerosols show a declining trend. 

\begin{figure}[h]
\centering 
\includegraphics[width=1\textwidth]{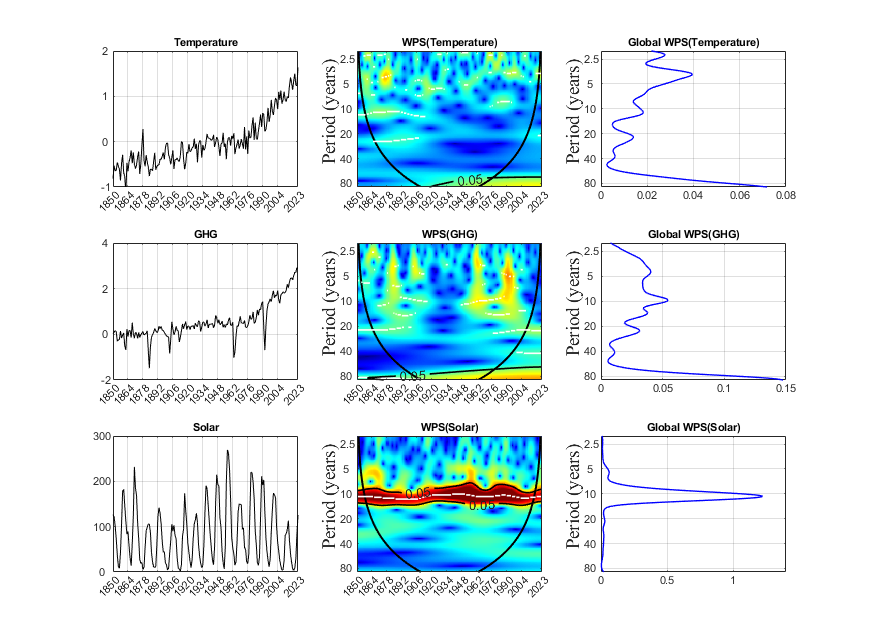}
\caption{\textit{(Left) Plot of each time series. (Middle) the corresponding wavelet power spectrum, for
cases: $Temperature$, $GHG$ and $Solar$. The black/grey contour designates the $5\%/10\%$ significance level. The cone of
influence,which is the region affected by edge effects, is indicated with a black line. The colour code for power ranges from blue
(low power) to red (high power). The white lines show local maxima of the wavelet power spectrum.(Right) Corresponding global wavelet power spectra.}}
\label{Figure: RR_1850_Temp_GHG_Solar_WPS}
\end{figure}

\subsection{Temperatures, GHG, and Sunspots}\label{Section: 3.2}
In the plots of the Wavelet Power Spectrum (WPS), ``cold'' areas are depicted in blue (indicating low volatility), while red indicates high volatility, with white lines showing local maxima of the WPS. Additionally, black (gray) contours represent the 5\% (10\%) significance level. The key insight from Figure \ref{Figure: RR_1850_Temp_GHG_Solar_WPS} is that both temperature and GHG show significant variations at longer frequencies, consistent with the presence of a common trend since the 1940s, with notable cycles at 5-10 years. The 10-11 year cycles are linked to solar radiation oscillations.

\begin{figure}
\centering 
\includegraphics[width=0.85\textwidth]{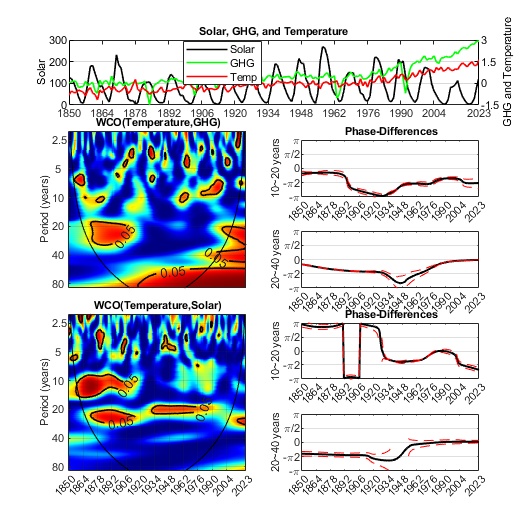}
\caption{\textit{(Top) Plot of time series. (Bottom Left) Wavelet Coherency (WCO).
Coherency Ranges form Blue (Low Coherency) to Red (High Coherency). (Bottom Right) Phase-Differences. The Black Line
Represents the Phase-Differences and the Red Lines the $95\%$ Confidence Level.}}
\label{Figure: RR_1850_Temp_GHG_Solar_WCO}
\end{figure}

Analyzing the bivariate relationships between temperatures and the different forcings in Figure \ref{Figure: RR_1850_Temp_GHG_Solar_WCO}, it shows intermittent coherency at shorter frequencies, with stable coherency at the longest frequency, again indicating a joint trend. Phase-difference diagrams show alternating in-phase and out-of-phase relationships -- it is interesting to note that for the lower frequency case, $\phi_{yx}$ is very close to zero, which confirms that temperatures and GHG are trending contemporaneously. 

As for the joint dynamics of temperatures and sunspots, the 10-11 year sunspot cycle is significant only early on, while a 20-year cycle dominates much of the 20\textsuperscript{th} century. Phase diagrams show alternating lead-lag relationships in the 10-20 year band, potentially due to complex feedback mechanisms, shifting to consistent temperature lagging in the 20-40 year band, aligning with the exogeneity of solar activity.

\begin{figure}[h]
\centering 
\includegraphics[width=.725\textwidth]{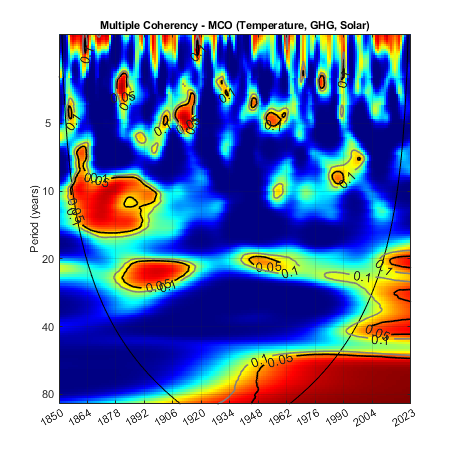}
\caption{\textit{Wavelet multiple coherency between $Temperature$ and forcings ($GHG$, $Solar$). The black/grey contour designates the $5\%/10\%$ significance level. The colour code for coherency ranges from blue (low coherency—close to zero) to red (high coherency—close to one).}}
\label{Figure: RR_1850_Temp_GHG_Solar_MCO}
\end{figure}

Precisely because the different forcings may interact with each other in potentially nonlinear forms and, in this way, further impacting temperatures, it is important to move from a simple bivariate analysis. Figure \ref{Figure: RR_1850_Temp_GHG_Solar_MCO} plots the multiple coherency (i.e., correlations across all time-frequency scales) of temperatures and the different forcings considered here. As also noticed for the bivariate plots, there is a significant and uninterrupted low frequency (above 50 years) coherency amongst the variables after the 1950's, whereas the heat map of Figure \ref{Figure: RR_1850_Temp_GHG_Solar_MCO} also suggests higher frequency cycles are important, but they are only occasionally significant.    

To gauge the interdependence between temperature and each of the forcings in the time-frequency domain, while accounting for the influence of other variables, we employ the notion of partial coherency (see Appendix). Indeed, it may be the case that if we observe a decrease in the (partial) coherency between temperature and a particular forcing in a specific region of the time-frequency space after controlling for a distinct forcing, we can infer that part of their interdependence stemmed from that third variable. Conversely, if the opposite occurs, we conclude that the third variable obscured the relationship. Given the heat map in Figure \ref{Figure: RR_1850_Temp_GHG_Solar_MCO}, we will focus on three frequency regions: 2-10 years to capture shorter cycles, 10-40 periods that should encompass the influence of solar activity on temperatures as seen in Figure \ref{Figure: RR_1850_Temp_GHG_Solar_WCO}, and longer cycles in the 40-80 years range, which will account for the trending behaviours.

\begin{figure}
\centering 
\includegraphics[width=1.049\textwidth]{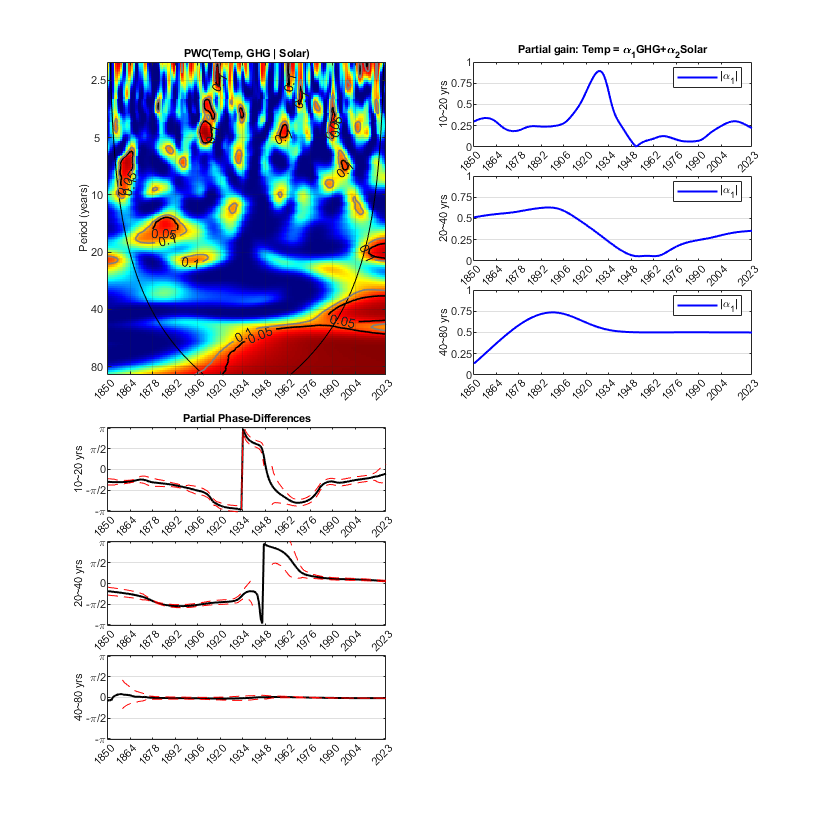}
\caption{\textit{(Top left) Partial wavelet coherency: $Temp$ vs. $GHG$ controlling for $Solar$. (Bottom left) Partial phase differences: $Temp$ vs. $GHG$ controlling for $Solar$. (Top right) Partial wavelet gain: $Temp$ vs. $GHG$ controlling for $Solar$.}}
\label{Figure: RR_1850_Temp_GHG_PWC}
\end{figure}

\begin{figure}
\centering 
\includegraphics[width=1.05\textwidth]{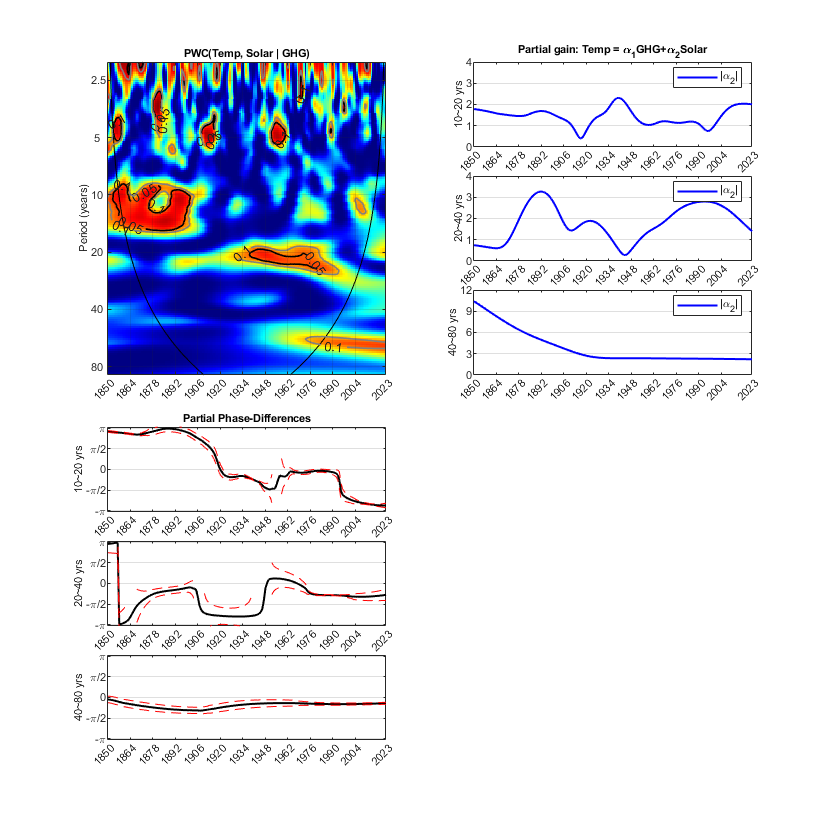}
\caption{\textit{(Top left) Partial wavelet coherency: $Temp$ vs. $Solar$ controlling for $GHG$. (Bottom left) Partial phase differences: $Temp$ vs. $Solar$ controlling for $GHG$. (Top right) Partial wavelet gain: $Temp$ vs. $Solar$ controlling for $GHG$.}}
\label{Figure: RR_1850_Temp_Solar_PWC}
\end{figure}

Analyzing the partial coherency between temperatures and GHG (Figure \ref{Figure: RR_1850_Temp_GHG_PWC}), controlling for sunspots, reveals that low-frequency cycles (40+ years) remain prominent post-1940s-1950s, indicating sunspots do not influence this relationship. The phase-difference diagrams show synchronous evolution at long-range frequencies, while shorter cycles exhibit switching correlations due to feedback effects. The partial gain panel indicates varying GHG effects over time (given by $|\alpha_1|$), with a surge at the 10-20 year and 20-40 year frequencies supporting the view of anthropogenic climate change driving temperature increases. The period of accelerated increases in temperatures, with a slight subsequent drop off, is consistent with the ``hiatus'' period identified in the literature (see \citealp{pretis_testing_2015} and \citealp{miller_dating_2020}), for example.\footnote{This spike in $|\alpha_1|$ is particularly salient in the results using CO$_2$ concentrations, reported in the Supplementary Appendix.}

Next, we consider the effect of solar activity on temperatures, while controlling for GHG. The partial coherency heat map in Figure \ref{Figure: RR_1850_Temp_Solar_PWC} shows that the effect of sunspots around the 20-year frequency is largely dissipated when compared to Figure \ref{Figure: RR_1850_Temp_GHG_Solar_WCO}. The partial gain figures show $|\alpha_3|$ declining for frequencies larger than 40 periods. In addition, the partial phase-differences diagrams confirm our earlier findings, i.e., temperatures are lagging most of the time, as solar radiance is exogenous.

\begin{figure}
\centering 
\includegraphics[width=1.05\textwidth]{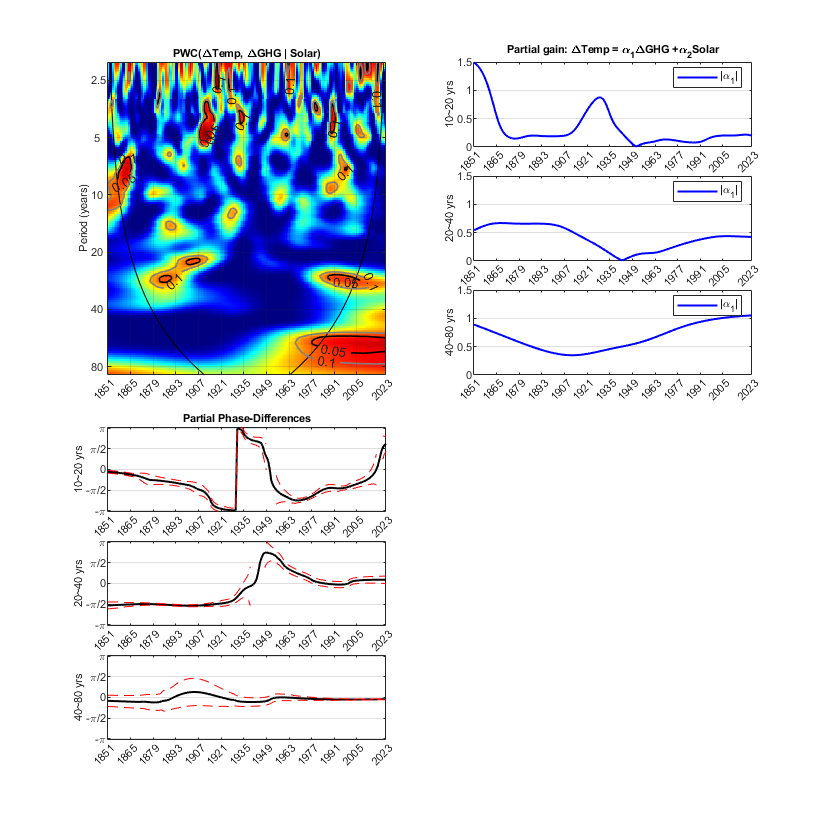}
\caption{\textit{(Top left) Partial wavelet coherency: $\Delta Temp$ vs. $\Delta GHG$ controlling for $Solar$. (Bottom left) Partial phase differences: $\Delta Temp$ vs. $\Delta GHG$ controlling for $Solar$. (Top right) Partial wavelet gain: $\Delta Temp$ vs. $\Delta GHG$ controlling for $Solar$.}}
\label{Figure: RR_1850_DTemp_DGHG_PWC}
\end{figure}

Finally, and for completeness, we also analyze the relationship between \textit{changes} in temperature (denoted $\Delta Temp$) and changes in GHG ($\Delta GHG$), controlling for sunspots. Naturally, with the variables thus transformed, the focus is on higher frequencies: Figure \ref{Figure: RR_1850_DTemp_DGHG_PWC} shows that most of the significant regions are for periods less than 20 years, but as in the previous case with the variables in levels, no systematic pattern emerges. For shorter cycles (between 10 and 20 years), there is some oscillation in the phase-difference diagram, also reflected in the fluctuations of the partial gain coefficient. Interestingly, for all frequencies, we observe the magnitude of the partial gain increasing, again suggesting an increasing effect of GHG in the acceleration of temperatures in the last few decades -- this is not unlike the results in \cite{Phellaetal2024}, for example.

\begin{figure}
\centering 
\hspace*{-1cm}  
\includegraphics[width=1.05\textwidth]{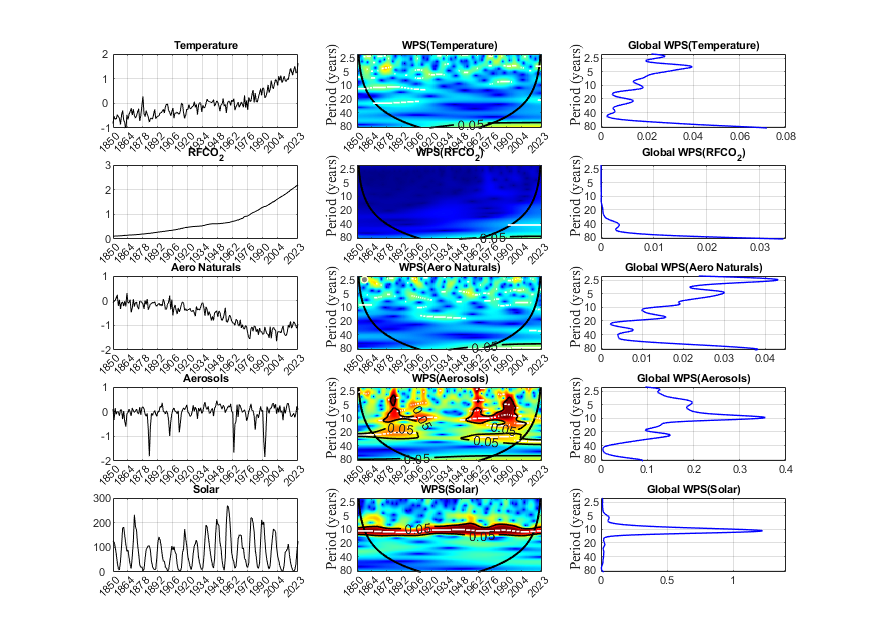}
\caption{\textit{(Left) Plot of each time series. (Middle) the corresponding wavelet power spectrum, for
cases: $Temperature$, $RFCO2$, $Aero Naturals$, $Aerosols$ and $Solar$. The black/grey contour designates the $5\%/10\%$ significance level. The cone of
influence, which is the region affected by edge effects, is indicated with a black line. The colour code for power ranges from blue
(low power) to red (high power). The white lines show local maxima of the wavelet power spectrum. (Right) Corresponding global wavelet power spectra.}}
\label{Figure: RR_1850_Temp_RFCO2_AN_Aero_Solar_WPS}
\end{figure}

\subsection{Decomposed Radiative Forcings}

In this section, we attempt to further understand how temperatures respond to natural and anthropogenic climate forces by considering the relationship between temperatures, CO$_2$ (the most abundant anthropogenic GHG), as well as natural and anthropogenic aerosols, while still controlling for solar forcing.\footnote{In a Supplementary Appendix, we include two additional analyses: in the first one, we use the same variables of this section and include CO$_2$ growth; the second exercise focuses on the relationship between CO$_2$ and temperatures with data starting in 1800.}

Figure \ref{Figure: RR_1850_Temp_RFCO2_AN_Aero_Solar_WPS} shows the wavelet power spectrum for the series of interest, including temperatures and solar which were previously presented. The profile for $RFCO_2$ is very similar to that of GHG as discussed for Figure \ref{Figure: RR_1850_Temp_GHG_Solar_WPS}, i.e., the salient feature is the trend captured by the lowest frequency cycle. Similarly to sunspots, the WPS of aerosols is dominated by the 10-year cycle, but this feature is not significant in the middle of the sample. In contrast, natural aerosols (denoted `AeroNaturals') are exclusively in the high (2.5-year) and low-frequency range given their downward trend.

\begin{figure}
\centering 
\includegraphics[width=.71\textwidth]{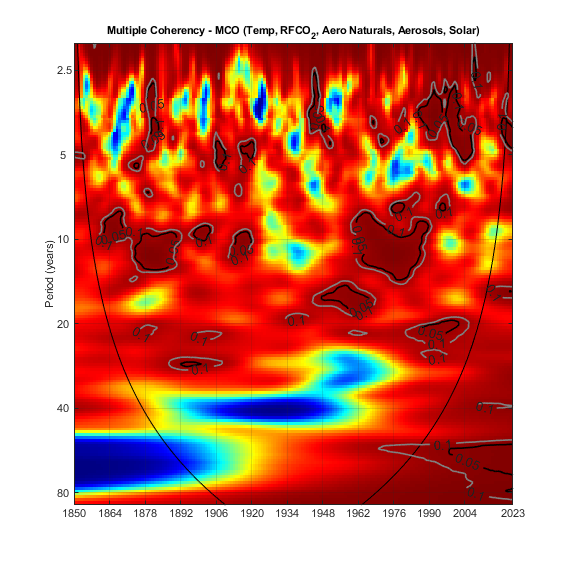}
\caption{\textit{Wavelet multiple coherency between $Temperature$ and forcings ($RFCO_2$, $Aero Naturals$, $Aerosols$ and $Solar$). The black/grey contour designates the $5\%/10\%$ significance level. The colour code for coherency ranges from blue (low coherency—close to zero) to red (high coherency—close to one).}}
\label{Figure: RR_1850_Temp_RFCO2_AN_Aero_Solar_MCO}
\end{figure}

On the other hand, the multiple coherency plot in Figure \ref{Figure: RR_1850_Temp_RFCO2_AN_Aero_Solar_MCO} reveals, aside from the 40-level frequency band, very strong dispersed patterns of correlation, both over time and across the different frequencies, with appreciably additional ``hot'' areas than in Figure \ref{Figure: RR_1850_Temp_GHG_Solar_MCO}. This means that $RFCO_2$, AeroNaturals and Aerosols share larger amounts of information with temperatures and solar than GHG alone does. A notorious example is the 10-year cycle during the second half of the twentieth century. 

\begin{figure}[H]
\centering 
\includegraphics[width=1.05\textwidth]{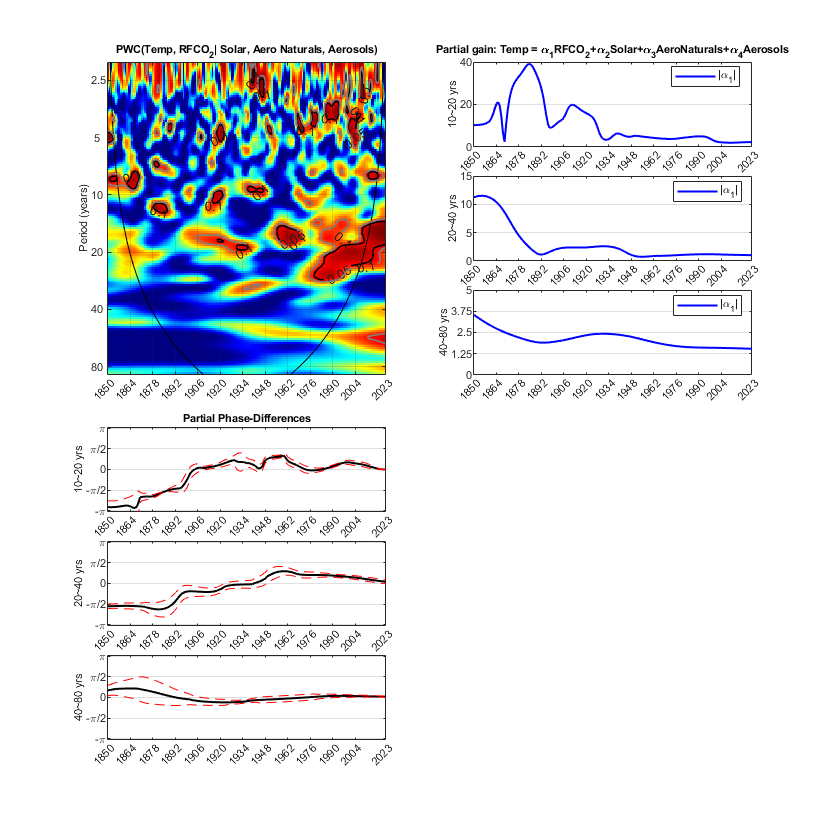}
\caption{\textit{(Top left) Partial wavelet coherency: $Temp$ vs. $RFCO_2$ controlling for $Solar$, $Aero Naturals$ and $Aerosols$. (Bottom left) Partial phase differences: $Temp$ vs. $RFCO_2$ controlling for $Solar$, $Aero Naturals$ and $Aerosols$. (Top right) Partial wavelet gain: $Temp$ vs. $RFCO_2$ controlling for $Solar$, $Aero Naturals$ and $Aerosols$.}}
\label{Figure: RR_1850_Temp_RFCO2_Solar_AN_Aero_PWC}
\end{figure}

\begin{figure}[H]
\centering 
\includegraphics[width=1.05\textwidth]{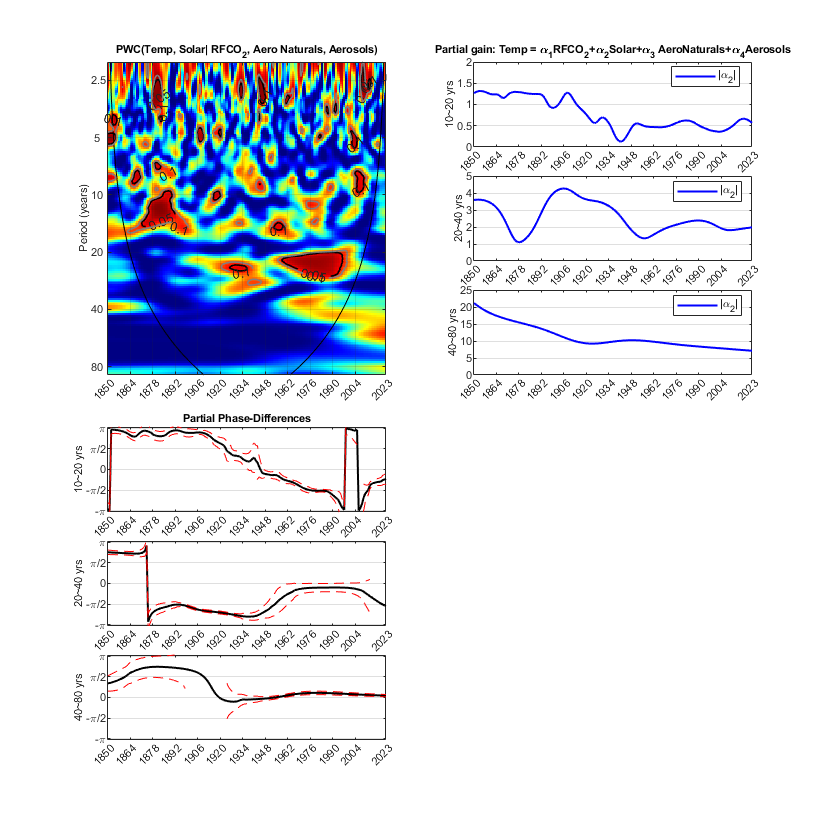}
\caption{\textit{(Top left) Partial wavelet coherency: $Temp$ vs. $Solar$ controlling for $RFCO_2$, $Aero Naturals$ and $Aerosols$. (Bottom left) Partial phase differences: $Temp$ vs. $Solar$ controlling for $RFCO_2$, $Aero Naturals$ and $Aerosols$. (Top right) Partial wavelet gain: $Temp$ vs. $Solar$ controlling for $RFCO_2$, $Aero Naturals$ and $Aerosols$.}}
\label{Figure: RR_1850_Temp_Solar_RFCO2_AN_Aero_PWC}
\end{figure}

\begin{figure}[h]
\centering 
\includegraphics[width=1.05\textwidth]{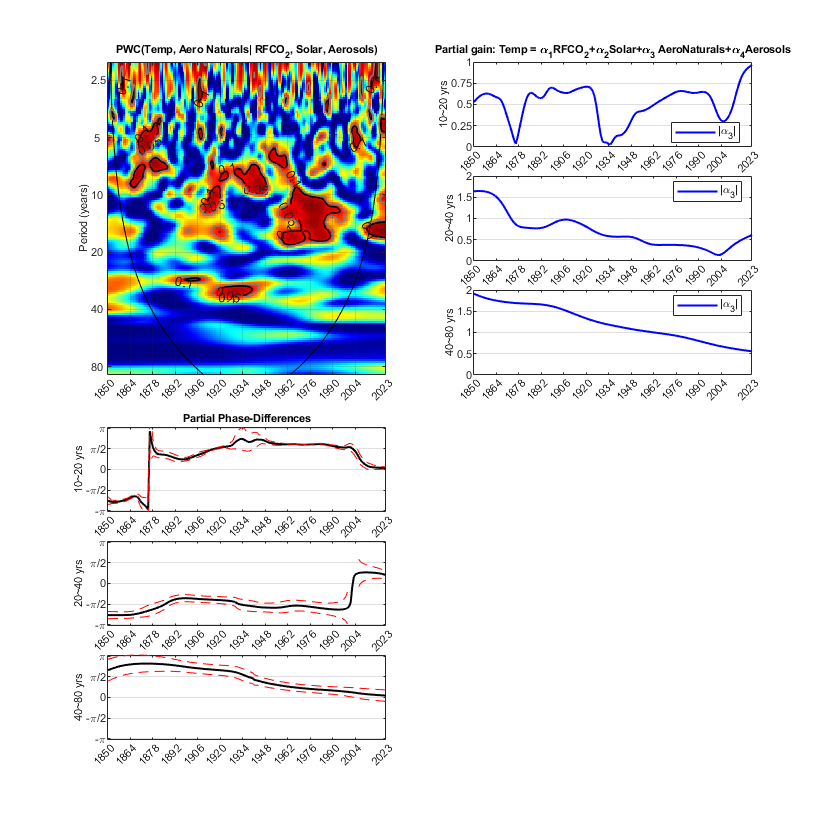}
\caption{\textit{(Top left) Partial wavelet coherency: $Temp$ vs. $Aero Naturals$ controlling for $RFCO_2$, $Solar$ and $Aerosols$. (Bottom left) Partial phase differences: $Temp$ vs. $Aero Naturals$ controlling for $RFCO_2$, $Solar$ and $Aerosols$. (Top right) Partial wavelet gain: $Temp$ vs. $Aero Naturals$ controlling for $RFCO_2$, $Solar$ and $Aerosols$.}}
\label{Figure: RR_1850_Temp_AN_RFCO2_Solar_Aero_PWC}
\end{figure}

\begin{figure}[h]
\centering 
\includegraphics[width=1.05\textwidth]{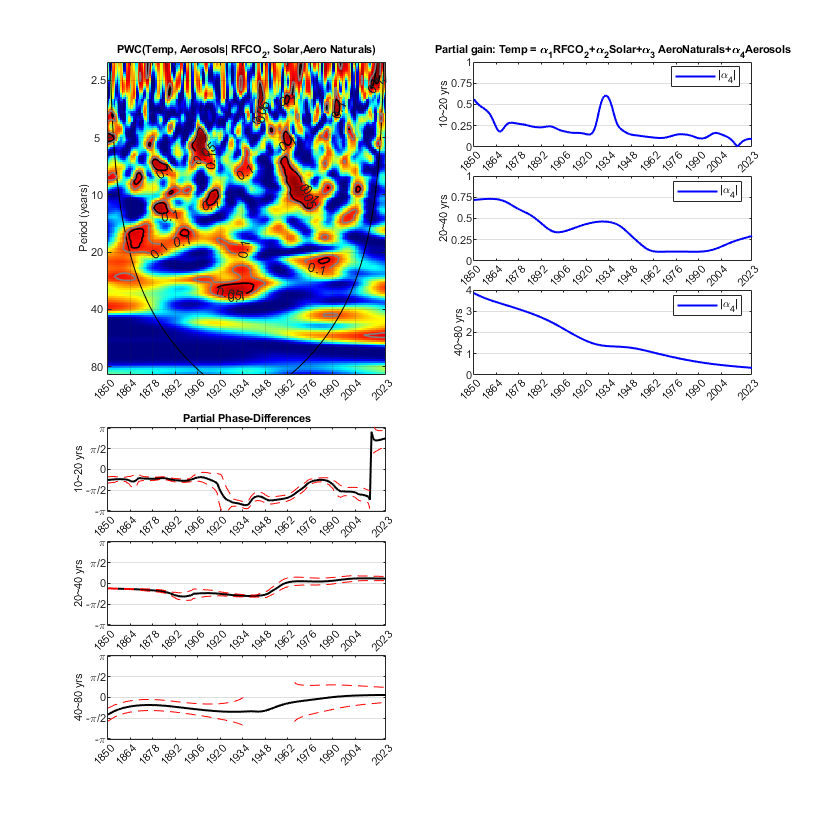}
\caption{\textit{(Top left) Partial wavelet coherency: $Temp$ vs. $Aerosols$ controlling for $RFCO_2$, $Solar$ and $Aero Naturals$. (Bottom left) Partial phase differences: $Temp$ vs. $Aerosols$ controlling for $RFCO_2$, $Solar$ and $Aero Naturals$. (Top right) Partial wavelet gain:  $Temp$ vs. $Aerosols$ controlling for $RFCO_2$, $Solar$ and $Aero Naturals$.}}
\label{Figure: RR_1850_Temp_Aero_RFCO2_Solar_AN_PWC}
\end{figure}

The (partialled out) contribution of each forcing is considered in Figures \ref{Figure: RR_1850_Temp_RFCO2_Solar_AN_Aero_PWC} to \ref{Figure: RR_1850_Temp_Aero_RFCO2_Solar_AN_PWC}. Regarding $RFCO_2$ (controlling for sunspots, aero naturals and aerosols), the results are comparable with those for GHG reported in Figure \ref{Figure: RR_1850_Temp_GHG_PWC} -- there is considerable contemporaneous correlation at lower ($40+$ years) frequencies as seen in the phase-difference plots, while the partial gain at shorter frequencies suggest that there is little interaction between temperatures and $RFCO_2$, especially since the beginning of the previous century. During this period, the contemporaneous correlation is also present for higher ($40-$ years) frequencies. For sunspots (see Figure \ref{Figure: RR_1850_Temp_Solar_RFCO2_AN_Aero_PWC}), the decomposition confirms earlier findings: the presence of a medium-run cycle of around 20 years and temperatures lagging most of the time. 

If in turn we consider the effects of natural aerosols (see Figure \ref{Figure: RR_1850_Temp_AN_RFCO2_Solar_Aero_PWC}), the procedure identifies significant cycles of around 5-20 years for a substantial part of the sample period, and a longer cycle of roughly 30-40 years, but circumscribed to the first half of the 20\textsuperscript{th} century. The partial gain coefficients suggest that for lower (20+ years) frequencies these associations decreased in importance over time. From the partial phase-differences, we note that contemporaneous correlation between temperatures and natural aerosols only occur at the end of sample period.

Finally, we study the relationship between temperatures and anthropogenic aerosols, after controlling for the other variables. From the partial coherency in Figure \ref{Figure: RR_1850_Temp_Aero_RFCO2_Solar_AN_PWC}, we observe several cycles, but scattered in both the time and frequency domains. Similar to previous cases, the partial gain decreased over time, namely for lower frequencies (40+ years). Also, the phase-diagram shows in-phase aerosols leading (temperatures lagging) for most of the time and frequencies. Only at the end of the sample we observe contemporaneous correlation at medium and lower (20+ years) frequencies. 

\bigskip

\bigskip

\section{Conclusion}

In this paper, we suggest that wavelet analysis can help researchers offer a more complete picture of the dynamics of climate change. Using a set of continuous wavelet tools, we study the relationship between temperatures and different (natural and anthropogenic) forcings, which allows us to observe changes over time, as well as heterogeneity across frequencies simultaneously. This is in contrast with most of the literature, which up to now has examined the relationship between temperatures and climate forcings either across time or frequency in isolation. Furthermore, the approach we employ is fully non-parametric and data-driven, imposing no underlying ``structural'' model, in contrast with the conventional practices in climate literature, therefore offering a complementary, statistical in nature, perspective. 

Our results show substantial variation along both the time and frequency dimension: sunspots are associated with temperatures at lower frequencies, namely with cycles of around 60-80, whereas changes in GHG or CO$_2$ are more relevant for shorter cyclical ranges, even shorter than 10 years. This observation underscores the importance of anthropogenic CO$_2$ emissions as a key driver of temperature variations, highlighting the critical role of human activities in shaping the Earth's climate system.

Of particular interest, but beyond the scope of this paper, is the relationship between temperatures and climate forcings across time, frequency, as well as geographical location. The documented geographical climate heterogeneity that exists with respect to both temperatures and CO$_2$ emissions (see \citealp{rivas_climate_2023}) brings to the forefront the need to incorporate such geographical differences into the models in order to be able to account for variations across all relevant domains. 

\bibliographystyle{authordate1}
\bibliography{Bibliography}

\newpage
\appendix

\section*{Appendix}\label{appmeth}

First, we provide a brief description of the wavelet tools used in the empirical analysis in section \ref{empanal} -- readers interested in the technical details of the continuous wavelet transform are referred to the surveys in \cite{Crowley2007}, \cite{GallegatiSemmler2014} and \cite{AguiarConrariaetal2014}. This is then followed by a description of the historical climate variables used in this study. 

\subsection*{A.1 Continuous Wavelet Transform}\label{cwt}

Researchers have employed Fourier transform and spectral analysis to assess the influence of frequencies in explaining the overall variance of a time-series $x_t$. However, Fourier analysis has a significant drawback: it lacks temporal information once the transformation is applied. As a result, while it allows identification of predominant cycles, it fails to indicate when these cycles are most significant. 

The Continuous Wavelet Transform (CWT), on the other hand, addresses this issue by translating $x_t$ into the time-frequency domain, essentially treating CWT as a function of two dimensions: time and frequency. Consequently, whilst the Fourier Power Spectrum identifies the primary cycles influencing the variance of a time-series, the Wavelet Power Spectrum (WPS) not only identifies these dominant cycles, but also specifies when they are most influential. 

The Discrete Wavelet Transform (DWT) provides a simple and efficient way of capturing the essential time-frequency features of a time series with a limited 
choice of wavelets, which explains its initial popularity in the literature. The CWT, while computationally more expensive, allows for more flexibility in the selection of wavelets and offers a more refined picture of the time-frequency patterns in the data. 

Consider the ``wave'' function $\psi(t)$, which is assumed to integrate to zero and also displaying a fast decay to zero. The rapid decay of $\psi$ suggests it behaves like a window function. Conversely, requiring its integral to be zero implies $\psi$ must exhibit oscillatory behaviour, allowing us to attribute a specific frequency to this function. From this ``mother wave'', we can generate wavelet offspring $\psi_{\tau,s}$ by scaling (by $s$) and translating by $\tau$ such that
\begin{equation}
\psi_{\tau,s}(t)=\frac{1}{\sqrt{|s|}}
\psi\left(\frac{t-\tau}{s}\right), \quad s,\tau \in R, s\neq 0.
\end{equation}
The scaling parameter $s$ controls the width of the wavelet, while the translation parameter $\tau$ controls the location of the wavelet along the $t$-axis. For $|s| > 1$, the windows $\psi_{\tau,s}$ become larger (hence, corresponding to functions with lower frequency), and for $|s| < 1$, the windows become narrower (hence, becoming functions with higher frequency).

Given a time series $x(t)$, its continuous wavelet transform with respect to the wavelet $\psi$ is a function of two variables, $W_x(\tau, s)$, given by
\begin{equation}\label{cwteq}
 W_x(\tau, s) = \int_{-\infty}^{\infty} x(t)\overline{\psi}_{\tau,s}(t) \, dt = \frac{1}{\sqrt{|s|}} \int_{-\infty}^{\infty} \overline{\psi}\left(\frac{t-\tau}{s}\right)dt, 
\end{equation}
where $W_x(\tau, s)$ represents the continuous wavelet transform and $\overline{\psi}_{\tau,s}(t)$ denotes the complex conjugate of the scaled and translated wavelet $\psi$ (with $\overline{(.)}$ denoting complex conjugate). As noted before, $s$ and $\tau$ control the width and the location along the t-axis of $\psi_{\tau,s}$.

Given that we are interested in analysing phase information in order to gauge lead/lag relationships in our climate series, it is necessary to select a complex-valued wavelet function. In what follows, we make use of the Morlet-type wavelet family, denoted as $\psi (t) = \pi - \frac{1}{4} e^{i6t} e^{-\frac{t^2}{2}}$, which has significant advantages. Specifically, when utilizing this wavelet, it becomes feasible to approximate the Fourier frequency $f$ as $f \approx \frac{1}{s}$, which makes interpretation straightforward. Given this correspondence between scale and frequency, we designate the $(t,s)$-plane as the time-frequency plane. As with the Fourier case, we can define the (local) wavelet power spectrum (WPS) of $x$:
\begin{equation}
WPS_x(t,s) = |W_x(t,s)|^2,
\end{equation}
which returns a measure of the distribution of $x$ in the time-frequency plane. Averaging the wavelet power over all times, we obtain the global wavelet power spectrum (GWPS), 
\begin{equation}
GWPS_x(s) = \int_{-\infty}^{\infty} |W_x(t,s)|^2 \, dt, 
\end{equation}
in all similar to the Fourier power spectrum. This is depicted in Figure \ref{fig:cwt} below.

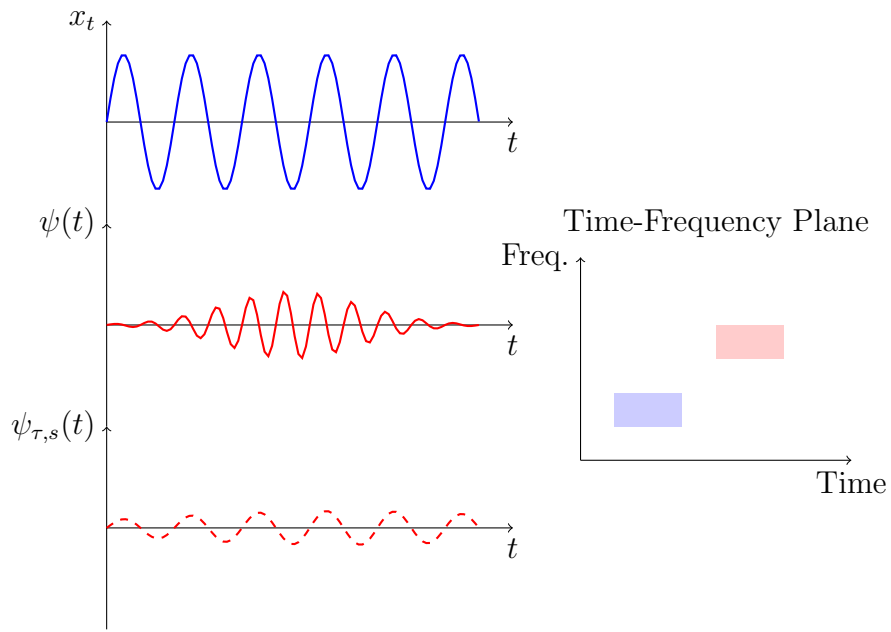
\begin{figure}[h]

\centering
\begin{tikzpicture}[scale=0.9]
    \draw[->] (0,5) -- (6,5) node[below] {$t$};
    \draw[->] (0,3.5) -- (0,6.5) node[left] {$x_t$};
    \draw[blue, thick] plot[domain=0:5.5, samples=100] (\x, {5 + sin(360*\x)});
    \node at (6.5, 4.5) {};

    \draw[->] (0,2) -- (6,2) node[below] {$t$};
    \draw[->] (0,0.5) -- (0,3.5) node[left] {$\psi(t)$};
    \draw[red, thick] plot[domain=0:5.5, samples=100] (\x, {2 + 0.5*sin(720*\x)*exp(-0.5*(\x-2.75)^2)});
    \node at (6.5, 1.5) {};

    \draw[->] (0,-1) -- (6,-1) node[below] {$t$};
    \draw[->] (0,-2.5) -- (0,0.5) node[left] {$\psi_{\tau,s}(t)$};
    \draw[red, dashed, thick] plot[domain=0:5.5, samples=100] (\x, {-1 + 0.25*sin(360*\x)*exp(-0.25*((\x-3.5)/2)^2)});
    \node at (6.5, -1.5) {};

    \draw[->] (7,0) -- (11,0) node[below] {Time};
    \draw[->] (7,0) -- (7,3) node[left] {Freq.};
    \fill[blue!20] (7.5,0.5) rectangle (8.5,1);
    \fill[red!20] (9,1.5) rectangle (10,2);
    \node at (9, 3.5) {Time-Frequency Plane};
\end{tikzpicture}
\caption{Continuous Wavelet Transform. The signal \( x_t \) (top) is analyzed using a mother wavelet \( \psi(t) \) (middle), which is scaled and shifted (bottom) to compute \( W(\tau,s) \), shown in the time-frequency plane (right).}
\label{fig:cwt}
\end{figure}

\subsection*{A.2 The bivariate case}\label{biv}

We now introduce several wavelet tools to investigate the relationship between two variables, $y$ and $x$, in the time-frequency domain. The complex wavelet coherency of $y$ and $x$ is defined as:
\begin{equation}
\rho_{yx} = \frac{S(W_{yx})}{\sqrt{S(|W_y|^2)S(|W_x|^2)}}
\end{equation}
where $W_{yx} = W_y\overline{W_x}$ represents the cross-wavelet power of $y$ and $x$, and $S$ denotes a smoothing operator in both time and scale. This complex wavelet coherency can be represented in polar form as $\rho_{yx} = |\rho_{yx}|e^{i\phi_{yx}}$, with $\phi_{yx} \in (-\pi, \pi]$. The modulus of $\rho_{yx}$ is referred to as the \textit{wavelet coherency}, representing the correlation between the two variables at each time and frequency. In turn, the angle $\phi_{yx}$ is termed the \textit{wavelet phase difference} between $y$ and $x$, providing insight into whether the two series are in-phase or out-of-phase, and also indicating the lead/lag relationship between $y$ and $x$. The corresponding interpretation is summarized in Table 1 in the text.

Moreover, \cite{MadlerScharnagl2014} introduced the concept of \textit{wavelet gain} between $y$ and $x$:
\begin{equation}
G_{yx}=\frac{|S(W_{yx})|}{\sqrt{S(|W_x|^2)}}
\end{equation}
The interpretation of $G_{yx}$ is similar to that of the Fourier gain of Engle (1976), i.e., we can view the wavelet gain as the regression coefficient of $y$ on $x$ (at each time and frequency), assuming no time lag between $y$ and $x$; when there is a lag, $G_{yx}$ can be interpreted as the regression coefficient if the series $x$ were appropriately shifted to eliminate any phase discrepancy, and $\phi_{yx}$ denotes the angle of that necessary shift.

\subsection*{A.3 Multivariate analysis}\label{multi}

The tools introduced in the preceding section can, with the appropriate reformulations, be extended to scenarios involving more than two series. Due to their complexity, we refrain from presenting the relevant expressions here and instead direct interested readers to \cite{AguiarConrariaetal2014}.
Given a series $y$ and $m$ series $x_i$ (where $i = 1, \ldots, m$), to evaluate the degree of linear time-frequency association between $y$ and the $m$ series $x_i$, one may compute the \textit{multiple wavelet coherency} between $y$ and $x_1, \ldots, x_m$. To assess the interdependence between variable $y$ and a specific variable $x_k$, while controlling for the effects of the other variables $x_j$ ($j = 1, \ldots, m; \ j \neq k$), we can employ the \textit{wavelet partial coherency}, the \textit{wavelet partial phase difference}, and the \textit{wavelet partial gain} between series $y$ and $x_k$, while controlling for $x_j$ ($j = 1, \ldots, m; \ j \neq k$). Specifically, the wavelet partial gain between $y$ and $x_k$, while controlling for the series $x_j$ ($j = 1, \ldots, m; \ j \neq k$), can be regarded as the coefficient of $x_k$ in the multiple regression of 
$y$  on the (appropriately shifted) $m$ variables $x_i$ ($i = 1, \ldots, m$).

In other words, the notions of multiple wavelet coherency and partial wavelet coherency can be seen as straightforward extensions of the respective concepts of multiple and partial correlation in the time-domain, or (Fourier) multiple coherency and partial coherency in the frequency domain, but now in terms of joint time-frequency analysis.

\end{document}